\documentclass[10pt,twocolumn,twoside]{IEEEtran}
\usepackage{nopageno}
\usepackage{amsmath,graphicx,cite}

\usepackage[mathcal]{euscript}
\usepackage{amssymb,amsmath}
\usepackage{algpseudocode}
\usepackage{verbatim}
\usepackage{graphicx}
\usepackage{mathtools}
\usepackage{tikz}
\usepackage{multirow} 
\usetikzlibrary{arrows,backgrounds}
\usepackage{amsmath,amstext,amsfonts,amssymb}
\usepackage{amsbsy}
\usepackage{amsthm}
\usepackage{diagbox}
\usepackage{mathabx}
\usepackage{color}
\usepackage{morefloats}
\usepackage[left=.98in,right=.98in,top=.75in]{geometry}
\usetikzlibrary{decorations.pathreplacing}

\definecolor{darkred}{rgb}{0.5,0.0,0.0}
\definecolor{darkblue}{rgb}{0.2,0.1,0.6}

\title{Task Oriented Channel State Information
Quantization}
\author{\IEEEauthorblockN{Hang Zou\IEEEauthorrefmark{1},
Chao Zhang\IEEEauthorrefmark{1}, and Samson Lasaulce\IEEEauthorrefmark{1}}

\IEEEauthorblockA{\IEEEauthorrefmark{1}LSS, CNRS-CentraleSupelec-Univ. Paris Sud, Gif-sur-Yvette, France}}

\begin{document}
%\ninept
%
\maketitle
\begin{abstract}
In this paper, we propose a new perspective for quantizing a signal and more specifically the channel state information (CSI). The proposed point of view is fully relevant for a receiver which has to send a quantized version of the channel state to the transmitter. Roughly, the key idea is that the receiver sends the right amount of information to the transmitter so that the latter be able to take its (resource allocation) decision. More formally, the decision task of the transmitter is to maximize a utility function $f(x;g)$ with respect to $x$ (e.g., a power allocation vector) given the knowledge of a quantized version of the function parameters $g$. We exhibit a special case of an energy-efficient  power control (PC) problem for which the optimal task oriented CSI quantizer (TOCQ) can be found analytically. For more general utility functions, we propose to use neural networks (NN) based learning. Simulations show that the compression rate obtained by adapting the feedback information rate to the function to be optimized may be significantly increased. 
\end{abstract}
\thispagestyle{empty}

%\begin{IEEEkeywords}
%Decisional Quantizer, Neural Network, energy efficiency
%\end{IEEEkeywords}

\maketitle
%==================================
%==================================

\section{Introduction\label{sec:Introduction}}

When a transmitter has to make a choice e.g., choosing a modulation coding scheme (MCS), quite often, it does not have the full knowledge of the state of the channel. A common scenario is that the receiver sends, through a feedback channel, a quantized version of the channel, hence the term "quantized CSI". Using quantized CSI has the advantage to limit the amount of signalling or overhead. When inspecting the literature or even standardization reports, general proposed quantization mechanisms are designed quite empirically. In this paper, our goal is to design a quantizer more formally. More specifically, we assume that the transmitter has a given radio resource allocation or control task to be performed. This task is translated as a function to be maximized. To perform this task, we want to know how the quantizer at the receiver should be designed, the motivation being that the nature of the resource allocation task should dictate the number of feedback resources to be used. To be more formal, if one assumes that the transmitter has to maximize a function under the form $u(x; g)$, $x$ being the (resource allocation) decision variable and $g$ the unknown parameters (namely, the channel state), one easily infer that using a quantized version of $g$ instead of $g$ will induce a loss of optimality. One important practical issue is to know more the relation between this loss and the number of resources dedicated to quantization. 

To show the relevance of our approach, we exhibit a case where the task oriented CSI quantizer (TOCQ) can be determined analytically. The case under consideration corresponds to a useful metric to measure energy-efficiency (EE) \cite{zappone}. The case study corresponds to a PC problem in which the transmitter has to choose a given discrete power level having a given feedback from the receiver(s). For the single-channel case (e.g., for one receiver or one band), the quantizer can be determined analytically. For the general case, we propose to use a NN approach to compute the quantizer. We provide numerical results which correspond to very recent results. These results are promising and should be developed further. 

%==================================
%==================================
\vspace{-0.8em}
\section{Problem Formulation\label{sec:Problem-Formulation}}

Assume the decision entity (a transmitter a priori) has to maximize a utility function $u: \mathcal{X}\times \mathcal{G} \rightarrow \mathbb{R}$, $\mathcal{X}$ being the decision space and $\mathcal{G}$ the parameter space. We assume that the effective decision space is discrete: $\mathcal{D}=\left\{ \boldsymbol{d}_{1},\dots,\boldsymbol{d}_{M}\right\} \subseteq\mathcal{X}$, $M < \infty$. This occurs for instance when the number of transmit power is discrete (consider e.g., some cellular communication standards or works such as \cite{gesbert} in which optimality is obtained with a finite number of decisions), the number of possible MCS is finite, or when selecting a band among a finite set (as in Wifi systems). A TOCQ $Q$ is defined by its cells used to partition the parameter space $\Pi =( \mathcal{C}_{1},\dots,\mathcal{C}_{M})$ and the mapping used to map each cell to a decision $\Phi : \left\{ \mathcal{C}_{1},\dots,\mathcal{C}_{M}  \right\} \rightarrow \mathcal{D} $. More precisely we want to obtain $Q^\star=(\Pi^\star, \Phi^\star)$ s.t. 
$Q^\star:\ \boldsymbol{x}\left(\boldsymbol{g}\right)=\boldsymbol{d}_{i},\ if\ \boldsymbol{g}\in\mathcal{C}_{i}$, with
${Q}^\star  \in\arg\max_{Q}\ \mathbb{E}_{\boldsymbol{g}}\left[u\left(\boldsymbol{x};\ \boldsymbol{g}\right)\mid Q,\boldsymbol{x}\in\mathcal{D}\right]$,
which also writes as
$\boldsymbol{x}^{*}\left(\boldsymbol{g}\right)=\boldsymbol{d}_{k},\ \textrm{if}\ \forall \ell\neq k,\ u\left(\boldsymbol{d}_{k};\ \boldsymbol{g}\right)\geq u\left(\boldsymbol{d}_{\ell};\ \boldsymbol{g}\right)$.  
\vspace{-0.6em}
%==================================
%==================================
\section{Application of TOCQ to power control}
\label{sec:application-to-PC}

\subsection{Analytical solution for a particular scenario of PC}

We assume that the utility function to be maximized at the transmitter is the following EE function (see e.g., \cite{zappone} \cite{belmega}):
$u^{\mathrm{EE}}\left(\boldsymbol{p};\boldsymbol{g}\right)=\frac{\sum_{i=1}^{N} f(\mathrm{SNR}_i)}{\sum_{i=1}^{N}p_{i}}$, where: $i\in \{1,2,...,N\}$ is an index which might represents the band, channel, or user index; $g_i>0$ is the gain of channel $i$; $\boldsymbol{p} = (p_1, ..., p_N)$ is the power allocation of vector; $\boldsymbol{g} = (g_1, ..., g_N)$ is the vector channels used by transmitter $i$; $\mathrm{SNR}_i$ is the signal-to-noise ratio associated with channel $i$ chosen as $\mathrm{SNR}_i = \frac{p_{i}g_{i}}{\sigma^{2}}$, where $\sigma^2$ is the received noise variance. Also, to obtain explicit quantities, we assume the efficiency function (which represents the packet success rate) to be $f(s) = \exp\left(- \frac{c}{s} \right)$ where $c \geq 0$ is a parameter related to spectral efficiency (see \cite{belmega}).  

Now, let us specialize to the single-channel case $N=1$ and denote by $ \mathcal{P} = \left\{ P_{1},\dots,P_{M}\right\} $ the set of possible transmit power levels. Then it can be proved that the optimal transition levels that define the intervals which partition the parameter space $\mathcal{G} = (0,\infty)$ are given by: 
$g_{0}^{*}\left(P_{i},P_{j}\right)=\frac{c\sigma^{2}\left[\frac{1}{P_{i}}-\frac{1}{P_{j}}\right]}{\ln P_{j}-\ln P_{i}}>0$, 
where $P_{i}<P_{j}$ are two consecutive power levels of $ \mathcal{P}$. Therefore, if the channel gain measured at the receiver $g$ belongs to the interval $[g_{0}^{*}\left(P_{i-1},P_{i}\right),  g_{0}^{*}\left(P_{i},P_{i+1}\right) ]$ then the receiver reports the label $i$ to the transmitter (which means that power level $P_i$ should be used). 

\vspace{-0.8em}
\subsection{A NN solution for general PC scenarios}

For $N>1$, providing a systematic analytical procedure to partition the parameter space for the EE power control problem constitutes a significant extension of this paper. In this paper, we propose to solve this problem by using a NN based learning procedure. This procedure can, in fact, be used for any utility function of the form $u(\textbf{p};\textbf{g})$. In the simulations, we have considered the above EE function and also the famous log-sum utility function: $u^{\mathrm{SR}}(\textbf{p};\textbf{g}) = \sum_{i=1}^N  \log(1+\mathrm{SNR}_i)$.    

\begin{figure}[tbh]
\begin{centering}
\includegraphics[scale=0.35]{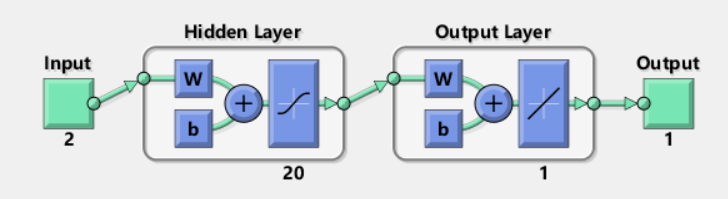}\caption{Feed-forward NN model for single-user $2-$band energy
efficiency problem \label{fig:Feed-forward-neural-network}}
\par\end{centering}
\end{figure}

A 3-layer \emph{feed-forward NN}, i.e., with only a hidden layer is chosen as our training model. The number of nodes for the input
layer, the hidden layer and the output layer is $2$, $20$ and $1$, respectively. The activation function for hidden layer is the sigmoid function and linear function for output layer. 
The structure of NN is illustrated in Fig.~\ref{fig:Feed-forward-neural-network}.  We can define our training
set as $\mathcal{T}_{n}\coloneqq\left\{ \boldsymbol{g}_{i},\ \theta_{i}\right\} _{i=1}^{n}$, 
where $\theta_{i}$ is the optimal decision label corresponding to
$\boldsymbol{g}_{i}$.
\vspace{-0.6em}
\vspace{-2em}

\subsection{Numerical results}
\begin{figure}[tbh]
\begin{centering}
\includegraphics[scale=0.35]{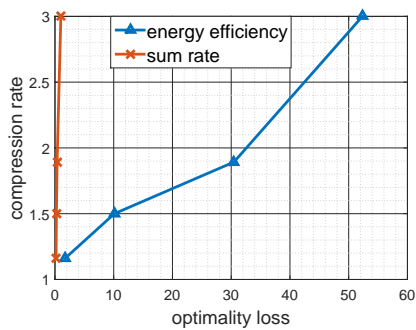}
\par\end{centering}
\caption{The compression rate in function of the optimality loss for single
user 2-band scenario for EE and sum-rate capacity. Compressing the channel gain for sum-rate function is easier than compressing the channel gain for EE function.}
\label{fig:compression_rate_optimality_loss}

\end{figure}

\vspace{-0.4em}
In this section, simulation results of a single user two bands scenario will be presented.  Define the optimality loss induced by quantization as
$\Delta u (\%)=\mathbb{E}_{\textbf{g}}\left[\left|\frac{u^{\star}\left( \textbf{g}\right)-{u}^{\mathrm{NN}}\left (\textbf{g}\right)}{u^{\star}\left( \textbf{g}\right)}\right|\right]\times 100$, where $u^{\star}\left( \textbf{g}\right)=\underset{\textbf{p}}{\arg\max}\,\,u(\textbf{p};\textbf{g})$ and  ${u}^{\mathrm{NN}}\left (\textbf{g}\right)$ is the performance derived by  our learning approach. Typical  model parameters are selected: $P_{\max}=5$ mW, $\sigma^2=1$mW and $c=1$. For all $i \in \{1,2\}$, the channel gain $g_{i}$ in band $i$ is assumed to be exponentially distributed, namely, its p.d.f. can be written as $\phi(g_i)=\exp(-g_i)$. In the NN model, $10000$ samples are divided into two parts: $9000$ training data and $1000$ test data.  The decision pair for EE is chosen
as following form $\left(0,P_{max}\frac{2i}{M}\right)$ or $\left(P_{max}\frac{2i}{M},0\right)$
for $\forall i\leq \frac{M}{2}$. As for sum-rate case, the decision
pair is uniformly chosen such that $P_{1}+P_{2}=P_{max}$. 
Define the compression rate $\gamma\left(\sigma\right)$ of a given
relative optimality loss $\sigma$ as
$\gamma\left(\sigma\right)=\frac{\log_{2}\left(M\left(1\%\right)\right)}{\log_{2}\left(M\left(\sigma\right)\right)}$
where $M\left(\sigma\right)$ is the required number of decisions such
that the loss $\sigma$ is met. Fig.~\ref{fig:compression_rate_optimality_loss}
represents the compression rate $\gamma$ in function of optimality loss
 for two bands in two cases.
\vspace{-1em}
\vspace{-0.7em}
In both  cases, the compression rate increases
as the optimality loss grows. For the EE maximization problem, the compression
rate decreases slowly while the optimality loss decreases and the loss
is always greater than $1\%$. As for the sum-rate maximization problem, the compression
rate declines rapidly while the optimality loss reduces and the optimality
loss is always less than $1\%$. In other words, in small optimality
loss regime, it is easier to compress the $\textbf{g}$ for the 
sum-rate problem than EE in two-band scenario. This can be explained by the fact that
 the explicit optimal decision function of sum-rate
, which is well known as the water-filling solution, is more concise
than the solution of EE.
\end{document}